\documentclass[final,5p,times,twocolumn,authoryear]{elsarticle}

\usepackage{amssymb}
\usepackage{graphicx}
\usepackage{amsmath}
\usepackage{natbib}
\usepackage{breqn}

\journal{}

\begin{document}

\begin{frontmatter}


 \ead{b.petkov@isac.cnr.it}

\title{Some considerations on the protection against the health hazards
associated with solar ultraviolet radiation}


\author{Boyan H Petkov}

\address{Italian National Research Council, Bologna, Italy}

\begin{abstract}
The present report briefly reviews the basic features of the current
strategy for the protection against the health harms caused by solar
ultraviolet (UV, $\sim$295 - 400 nm). The emphasis has been made
upon the erythema as being the best-studied UV harm and the ability
of UV irradiance to damage the deoxyribonucleic acid (DNA)
molecules, which leads to carcinogenesis. The erythemally weighted
UV irradiance that determines the ultraviolet index (UVI),
represents a common measure of the solar UV radiation level at the
Earth's surface and the current protective messages have been made
by using UVI as a basic parameter. However, such an approach seems
insufficiently grounded to be used also in the messages against the
skin cancer, bearing in mind the different nature of the erythema
and DNA lesions. In this context, an enlargement of the strategy
basis by including additional biological effects of UV radiation
studied during the past years has been discussed. For instance, the
weight of the spectral UV-A (315 - 400 nm) band that in practice had
been neglected by UVI definition can be enhanced since it was found
to play an important role in DNA damaging. In addition, features of
the contemporaneous life style can be taken into account together
with some people habits.
\end{abstract}



\begin{keyword}
Solar UV irradiance \sep Skin cancer \sep Erythema \sep Protection
against UV radiation




\end{keyword}

\end{frontmatter}


\section{Introduction}
\label{Intr} Research performed over the last century recognized the
solar ultraviolet (UV) radiation reaching the ground ($\sim 295 -
400$ nm) as an environmental factor that is able to cause a variety
of damage in the human organism like erythema, cataract and skin
aging
\citep{Ambach_Blumthaler1993,Mullenders2018,WHO2002,WHO2003,Debacq-Chainiaux2012}.
In addition, the exposure to solar UV irradiance can trigger
carcinogenesis
\citep{Elwood_Jopson1997,Jhappan2003,Tucker2008,Green1985,Karran_Brem2016,Ambach_Blumthaler1993}.
On the other hand, UV irradiance stimulates the production of
vitamin D, which enhances the resistance to various diseases,
including cancer
\citep{Reichrath2014,McKenzie2009,WHO2003,Passeron2019}. Thus, the
exposure to solar UV radiation should not be completely avoided, but
a balance between damage and benefit need to be sought. All of these
circumstances led to the elaboration of rules assuring safety
exposure to solar UV radiation and production of supplements, like
sunscreen creams that improve an organism's defence
\citep{WHO2002,WHO2003,Harrison2016}. These preventions formed a
protective strategy that was aimed at reducing the harmful effects
of the solar UV radiation. An important requirement for that is the
knowledge about UV irradiance at the surface and its diurnal and
seasonal variations. Currently, various satellite devices
\citep{Tanskanen2006,Schmalwieser2017} and ground-based instruments
provide both operative information about the distribution of UV
radiation and data for studying its behaviour
\citep{Bernhard2015,Bais2015,Petkov2014,Seckmeyer2012}.

However, despite the precautions to reduce the health hazards
associated with solar UV radiation, statistics generally shows that
the incidence of skin cancer has increased during the past decades
although the improvements in the diagnostic approaches
\citep{Tucker2008,Apalla2017,Corona1996,Lai2018,WHO2003}.
\cite{Guy2015} reported a doubling of the melanoma incidence rate in
the United States for the period from 1982 to 2011.
\cite{Wallingford2015} found a nearly 2\% increase of melanoma
occurrences per year for young English people (at the age between 14
and 24 years). Meantime, the 6-12\% decrease of Australian youth was
assumed to be a result of the protective strategies. The cases of
basal cell carcinoma (BCC) that is the most spread malignancy in
Caucasian population were found to increase with a rate of about 2\%
per year in Australia followed by 0.4\% per year in the United
States and nearly 0.15\% per year in Europe
\citep{Lai2018,Apalla2017}. According to other estimates the BCC
occurrences in the United States rose from 20\% to 80\% during the
period between the mid 1970 and 2006, while the incidence of
squamous cell carcinoma (SCC) increased with a rate of 3-10\% per
year for the same period \citep{Wadhera2006}. The number of SCC
cases duplicated from 2004 to 2012 in Belgium \citep{Callens2016}
and a doubling for man and a tripling for women were registered in
Denmark between 1978 and 2007 \citep{Birch-Johansen2010}.

Such a brief extract from the statistics of the skin cancer
occurrences lead to the impression of gaps in the current protective
strategy against the solar UV radiation since the exposure to UV
irradiance was indicated as the main factor for increasing cancer
incidence by most of the authors who provided the statistical data.
The next sections discuss the basis of the present protective scheme
and some new achievements in the researches related to the UV
radiation effects in the organisms. An emphasis was made on the
impact of UV radiation on the structure of the deoxyribonucleic acid
(DNA) molecule and assumptions about an enlargement of the
protective strategy basis have also been discussed.

\section{Solar UV irradiance at the ground and its impact on the human organism.}
\label{Sec2} Being the main energy source for the Earth, the solar
radiation is one of the most important environmental factors that
drives various chemical and dynamical processes in the atmosphere,
hydrosphere and biosphere \citep{Melnikova_Vasilyev}. Emitted by the
Sun electromagnetic radiation pertains to a large spectral range
starting from X-rays ($10^{-10}$ m) and reaching to the radio waves
$10^{-1}$ m \citep{Brasseur_Solomon}. UV irradiance occupies the
range from 100 nm to 400 nm (from $1\cdot10^{-7}$ to $4\cdot10^{-7}$
m) and in turn, it is subdivided into UV-C (100 - 280 nm), UV-B (280
- 315 nm) and UV-A (315 - 400 nm) spectral bands \citep{WHO2002}.
The radiation pertaining to the 400 - 700 nm wavelength interval,
which can be felt by the human eye composes the visible spectral
band, while the wavelengths from 700 nm to $10^{6}$ nm are named as
infrared radiation. The last two spectral ranges together with UV
band contain about 99\% of the solar irradiance reaching the Earth
surface.

\begin{figure}\centering
  \includegraphics{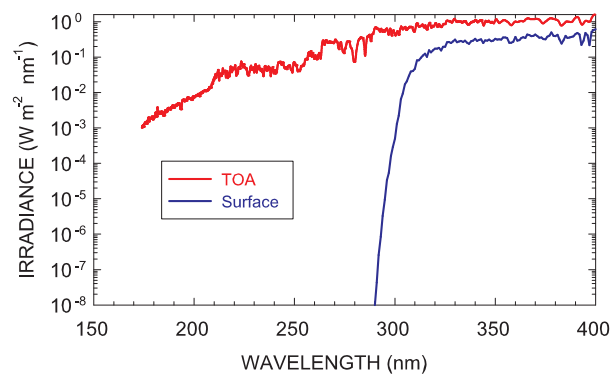}
 \caption{A fragment of the solar spectral irradiance at the top of the atmosphere (TOA)
  and an example of irradiance at the surface level presented by measurements in Bologna,
  Italy on 1 August 2006,   09:45 UTC. The TOA spectrum was taken from the library of the Tropospheric
  Ultraviolet-Visible (TUV) radiative transfer model \citep{Madronich_Flocke1997}.}\label{Fig1}
\end{figure}

Figure \ref{Fig1} presents the spectral distribution of the solar UV
radiation entering the terrestrial atmosphere and an example of the
typical spectral radiation reaching the Earth's surface. As can be
seen the wavelengths below about 295 nm turn out to be strongly
attenuated in the atmosphere that is due to the absorption mainly by
the oxygen (until about 250 nm) and ozone in the stratosphere
\citep{Brasseur_Solomon}. Figure \ref{Fig1} shows also that UV-A
solar irradiance at the ground is more intensive than UV-B by about
an order of magnitude.

UV radiation possesses enough energy to affect bioprocesses in the
human's organism, focusing its impact on the skin and eyes that
realize contact with the environment. The upper skin layer, the
stratum corneum that is built by dead keratinocytes absorbs
significantly in UV-B spectral range \citep{Bruls1984} reducing its
penetration into the inner epidermis layers. It should be mentioned
that the thickness of the stratum corneum could change depending on
the skin hydration and solar UV irradiation
\citep{Egawa2007,Bouwstra2003, Bruls1984}. Nevertheless, a certain
amount of UV-B together with UV-A radiation can reach the internal
epidermis strata causing various damage effects, such as erythema,
skin aging, photodermatosis and are able to trigger carcinogenic
processes in the human skin
\citep{Ambach_Blumthaler1993,Yu_Lee2017,Elwood_Jopson1997,Jhappan2003,Setlow1993,Boukamp2005,WHO2003}.

The erythema, which is one of the more frequently occurring skin
damages is an inflammatory dermatological effect that appears
several hours after the exposure to the UV radiation and it is
characterised by red skin due to the enhanced blood flow in the
stricken area
\citep{Agache1998,Flarer1955,Nishigori2014,Roy_Gies2017}. In light
skin people (skin type II-IV, \cite{WHO2002}), such a condition
subsides within several days passing to the tanning that is
determined by enhanced melanin concentration in the upper epidermis
layers \citep{D'Mello2016,Baker2017,Maddodi2012,Kaur_Saraf2011}. It
is usually considered that this enhancement is due to induced
production of melanin by melanocytes, while recent researches
indicated that solar UV irradiance stimulates its transport from the
basal layer \citep{Tadokoro2005}. Together with the stratum corneum,
the melanin is an additional epidermal shield that reduces the
propagation of UV irradiance into the deeper skin layers because of
its absorption in this spectral range
\citep{Anderson_Parrish1981,Brenner_Hearing2008,Solano2014,Premi2015}.

UV radiation is able to impact also the sub-cellular structures such
as the DNA molecules causing an array of genomic mutations
\citep{Ravanat2001,Mullenders2018} that can lead to cancer
formation. The human organism possesses various mechanisms able to
repair the lesions in DNA caused by UV irradiance and the efficiency
of these mechanisms is very important for the human health
\citep{Sinha2002,Rastogi2010,Mouret2008,Smith2000,Bustamante2019}.
It was established that the cell's inability to repair the DNA
damages is one of the factors that could lead to carcinogenesis
\citep{Jhappan2003,Lehmann2011,Yu_Lee2017}.

\begin{figure}
  \includegraphics{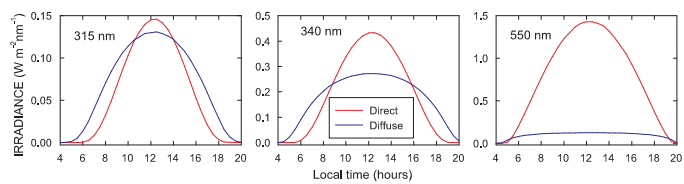}\\
  \caption{Direct and diffuse components of the solar irradiance
  at different wavelengths indicated in the corresponding panels. The curves were computed by
  the TUV radiative transfer model for Bologna, Italy on 21 May 2009 and illustrate the solar
  irradiance in case of clear sky conditions. The presence of clouds around the Sun, without
  covering it could enhances the diffuse component.}\label{Fig2}
\end{figure}

The propagation of the solar UV irradiance through the atmosphere is
characterised by enhanced scattering from the gas molecules
(Rayleigh scattering) that is much more intensive for the shorter
wavelengths. As a result, the diffuse component of the solar UV
irradiance at the ground turns out to be higher than the
corresponding components in the visible and infrared bands. Figure
\ref{Fig2} shows time-patterns of the direct and diffuse solar
irradiance at three wavelengths for clear sky conditions as were
evaluated by the Tropospheric Ultraviolet-Visible (TUV) radiative
transfer model \citep{Madronich_Flocke1997}. It is seen that the
diffuse component is almost as intensive as the direct one for UV-B
wavelengths, which relation gradually changes towards longer
wavelengths. The diffuse radiation becomes nearly half as much than
direct for UV-A and an order of magnitude less intensive for visible
and infrared (the case not shown in the figure) spectral bands.
Bearing in mind these features it can be concluded that an efficient
protection against the visible or infrared solar radiations would be
achieved by reducing the corresponding direct components, or staying
in shadow, for instance. However, this is not a reliable protection
against the UV irradiance since the diffuse component that could
arrive from any direction is not completely stopped. In addition, it
should be pointed out that we cannot feel the level of solar UV
irradiance as we could have an idea about intensity of the visible
and infrared radiation through the eyes and sensation of heat by the
skin, respectively. These circumstances would not allow us to have a
foreboding of the peril in some specific conditions like those in
high mountain, above about 2000 m, where the coming UV irradiance is
higher than at sea level \citep{Siani2008}. Additional forcing
factors could be a surrounding of snow covered surfaces and the
presence of haze that reduces the visible irradiance, making an
impression of low irradiation. Actually, such environment can
contribute to appreciable enhancement of the diffuse UV irradiance
since the snow reflects more than 70\% of UV and the haze,
especially if it is composed by very small particles could
additionally scatter the light according to the Rayleigh law. These
conditions are very dangerous for the skin and the eyes since they
lead to erythema and/or photokeratitis. Similar damage could occur
after a long staying under an umbrella on the beach, where the umbra
creates the feeling of protection. However, the elevated reflection
capacity of the sand in UV band (about 20\%) together with the
reflection from the water and Rayleigh scattering enhance the
diffuse component of the solar UV radiation.

Above extreme cases, as well as the usual everyday UV exposure imply
the knowledge of the UV irradiance level as an important requirement
for effective protection. For that reason continuous observations of
solar UV radiation have been developed within the past few decades
and nowadays the information about the solar UV irradiance is
available for almost all over the world
\citep{Tanskanen2006,Schmalwieser2017,Mims2019}.

\section{Basic features of the present solar UV protective strategy.}
\label{Sec3} Current messaging around sun protection to avoid the
harmful effects of exposure to solar radiation requires the
monitoring of the solar irradiance at the Earth's surface. Such
activity allows the estimations of the solar energy (dose) absorbed
by a unit horizontal surface and, as a result, the elaboration of
recommendations about exposure to solar irradiance. Since the
photobiological effects are spectral sensitive, the dose, related to
a certain damage has its specific meaning that is the subject of the
next subsection.

\subsection{Biologically active UV doses and estimation of the irradiation times.}
\label{Sec3.1} The effects produced by solar UV irradiation in human
organism depend on the wavelength $\lambda$, so that the radiation
at different wavelengths has different capacity to produce a certain
effect and the function determining this ability is named as action
spectrum $A_{eff}(\lambda)$. Weighting the spectral solar irradiance
$I(\lambda,t)$ at time $t$ by the action spectrum we obtain the
radiation $I_{eff}(t)$, able to produce the considered effect in
humans:

\begin{dmath}\label{Eq1}
    I_{eff}(t)=\int_{280 nm}^{400 nm}I^{'}_{eff}(\lambda,t)\,\text{d}\lambda= \\
    \int_{280 nm}^{400 nm}
    I(\lambda,t)\,\,A_{eff}(\lambda)\,\,\text{d}\lambda\,.
\end{dmath}

\begin{figure}\centering
  \includegraphics{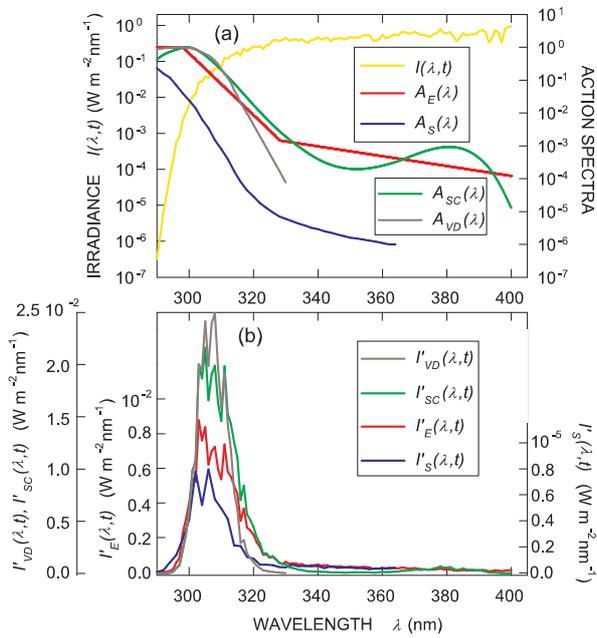}\\
  \caption{Upper panel (a) presents the solar spectral UV irradiance $I(\lambda,t)$ as was measured
  in Bologna, Italy on 21 May 2009, 12:00 UTC together with the erythema ($A_{E}(\lambda)$),
  Setlow DNA damage ($A_{S}(\lambda)$), skin cancer ($A_{SC}(\lambda)$) and vitamin D formation
  ($A_{VD}(\lambda)$) action. The lower panel (b) exhibits
  the effect of irradiance  weighting expressed by the irradiances $I^{'}_{eff}(\lambda,t)$ }\label{Fig3}
\end{figure}

Figure \ref{Fig3}(a) exhibits the action spectra of the (i) erythema
$A_{E}(\lambda)$ \citep{McKinlay_Diffey1987}; (ii) DNA damage
$A_{S}(\lambda)$ \citep{Setlow1974}; (iii) skin cancer in albino
hairless mice $A_{SC}(\lambda)$, which was corrected for human skin
\citep{Gruijl1993}; and (iv) vitamin D action spectrum
$A_{VD}(\lambda)$ \citep{MacLaughlin1982}. Figure \ref{Fig3}(b)
shows the irradiances
$I^{'}_{eff}(\lambda,t)=I(\lambda,t)A_{eff}(\lambda)$ that indicate
the wavelengths contributing corresponding $I_{eff}(t)$. It can be
seen that the UV-B band presents a dominant role in all selected
weighted irradiances. In fact, as Fig. \ref{Fig3} indicates the
weight of the UV-A band determined by the action spectra is even
lower than 2 order of magnitude.

For the purposes of the protective messages the dimensionless UV
index (UVI) was determined as $\text{UVI}(t) = I_{E}(t)/(0.025\,\,
\text{Wm}^{-2})$ \citep{WHO2002,Lucas2018}.

The solar energy associated with a certain effect and absorbed
during the time interval $\Delta T_{eff}$, or the corresponding dose
$D_{eff}$ can be evaluated as:

\begin{equation}\label{Eq2}
    D_{eff}(\Delta T_{eff})=\int_{\Delta
    T_{eff}}I_{eff}(t)\,\text{d}t\,.
\end{equation}
In a particular case when the action spectrum is
$A_{UV-B}=(1,\,\,\text{for}\,\, 280 \leq
\lambda\leq315\,\,\text{nm}\,\,\text{and}\,\, 0,\,\,\text{for}\,\,
315 <\lambda\leq 400\,\,\text{nm})$ or
$A_{UV-A}=(0,\,\,\text{for}\,\,280\leq \lambda\leq315\,\,\text{nm}
\,\,\text{and} \\ \,\, 1,\,\,\text{for}\,\, 315 <\lambda\leq
400\,\,\text{nm})$, Eqs. (\ref{Eq1}) and (\ref{Eq2}) give the UV-B
and UV-A irradiances and doses, respectively.

If we know the solar irradiance coming on the skin together with the
action spectrum $A_{eff}(\lambda)$ and the minimal dose
$D_{eff,min}$ that is able to cause a certain effect, the Eqs.
(\ref{Eq1}) and (\ref{Eq2}) allow the estimation of the exposure
time $\Delta T_{eff,min}$ needed to achieve the dose $D_{eff,min}$.
In case of damage effect the irradiation above the time $\Delta
T_{eff,min}$ enhances the probability of the damage occurrence and
this time appears to be a parameter quite appropriate for protective
messages.

Among all the known damages caused by UV radiation, the erythema,
which has been examined from the beginning of the last century
\citep{Coblentz_Stair1934,Luckiesh1930,Anders1995,Dornelles2004}
turns out to be the deepest studied. As a result, the erythemal
action spectrum $A_{E}(\lambda)$ together with the threshold dose
$D_{E,min}$, named as minimal erythema dose (MED) have been reliably
determined. Bearing in mind the different sensitivities of the
people to UV radiation, six types of skin were determined
\citep{Fitzpatrick1988,Roy_Gies2017}. Measurements of the
corresponding MED \citep{Dornelles2004,Bieliauskiene2019} allowed
the estimation of the exposure period $\Delta T_{E,min}$ within
which the erythema can be avoided \citep{McKenzie2009}. Actually,
such an estimation can be currently made only for the erythema
because $A_{eff}(\lambda)$ and $D_{eff,min}$ for other damages
caused by solar UV radiation have not been adopted yet. While the
erythema was studied on voluntaries, the examinations of the DNA
damages were made on animals or through in vitro experiments that
creates some restrictions for applying them to humans
\citep{Setlow1974,Gruijl1993,He2006,Wischermann2008}. For that
reason, the erythemal solar radiation, especially UVI has become the
basic parameter for the protective strategies.

\subsection{Protection against the damages induced by solar UV radiation.}
\label{Sec3.2} The estimation of irradiation time allows the
elaboration of protective messages that suggest some rules about the
exposure to solar light depending on the current UVI
\citep{WHO2002}. Such rules, aimed to avoid erythema include the
duration of the exposure for different skin types, depending on the
season and geographical location, recommendations about the clothes,
etc. In case of need, one could estimate the individual irradiation
time for both erythema damage and vitamin D production knowing his
skin type, solar irradiance and the sun protection factor (SPF) of
the sunscreen used \citep{Antoniou2008,McKenzie2009,Passeron2019}. A
method for more precise estimation of the doses that takes into
account the specific shape of the human body was also elaborated
\citep{Seckmeyer2013} together with empirical approaches connecting
the doses really absorbed by the skin with the doses on horizontal
surface that are usually provided by the solar irradiance
instruments \citep{Siani2008}.

The tanned fair skin assumes an enhanced protective capacity of the
organism despite that it can assure only a limited protection
corresponding to SPF of no more than 4 \citep{WHO2002,Wood1999}.
Hence, the caution about solar UV exposure should not be ignored
after tanning.

Generally speaking, above the description presents the core of the
current protective strategy against the damages associated with the
solar UV radiation. It was constructed on the basis of erythema,
since it uses UVI and $D_{E,min}$ for protective messages. However,
an important question can be asked: Is this strategy really able to
provide as reliable defence against the possible triggering of skin
cancer as the defence against the erythema? Indeed, the DNA
absorption spectrum is very similar to the erythemal action spectrum
\citep{Kiefer2007} but this is not a sufficient ground for a
positive answer. The erythema is an immediate effect of the UV
irradiation and it was assumed that the risk of long-term cancer
formation decreases if the sunburning can be avoided
\citep{Wood_Roy2017,Nishigori2014,Green1985,Elwood_Jopson1997}. Such
assumption was based mainly on epidemiological studies and it could
hardly give any quantitative relationships, which in turn would
allow the elaboration of concrete rules as in the case of erythema.

The progress in the research about the ability of UV irradiance to
trigger carcinogenesis could be used to improve the protective
strategy against the harms of solar radiation. Following the current
approaches such improvement could be achieved by introducing
additional parameters that describe the recently examined effects
and by assessing their impact on exposure time. Similarly to the
role of the skin types for erythema the parametrization of a certain
effect requires a common classification of the main individual
features that would allow to launch protective messages to public.
In other words, the basis of the strategy should be enlarged
considering not only erythema.

\section{Additional effects that could be used for protective messages.}
\label{Sec4} As pointed out in the previous section, the current
protective strategy concerns in practice only UV-B solar radiation.
However, despite that the absorption of DNA molecule in UV-A band is
negligible the latest researches clearly showed that UV-A irradiance
is an important factor for carcinogenesis as well \citep{Agar2004}.
This is mainly due to its ability to affect DNA indirectly through
its capacity to induce the generation of reactive oxygen species
(ROS). Such oxidative stress damages the DNA nucleobases and
inhibits the nucleotide excision repair that restores the DNA
lesions caused by UV-B irradiance
\citep{Karran_Brem2016,McAdam2016,Ravanat2001,Mullenders2018,Jhappan2003,Delinasios2018}.
In addition, \cite{He2006} found that chronic irradiation of
keratinocytes by UV-A irradiance can induce apoptotic resistance and
as a result to enhance the probability of carcinogenesis. Despite
that UV-A radiation was considered to cause predominantly non
melanoma cancer, last studies revealed pathways also for melanoma
carcinogenesis \citep{Premi2015}.

\begin{figure}
  \includegraphics{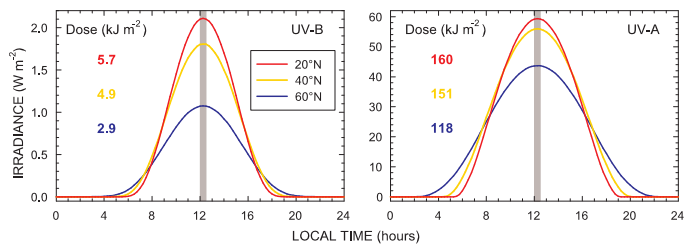}\\
  \caption{Time patterns of the solar UV-B (left) and UV-A (right) radiations coming
  to a horizontal area of the Earth surface at $20^{o}$N, $40^{o}$N and
  $60^{o}$N, respectively on 30 June 2019. The corresponding doses absorbed by a
  horizontal surface and evaluated for a half-hour
  interval between 12:00 and 12:30 local time (shadowed areas) are also indicated in the
  graphs.}\label{Fig4}
\end{figure}

Recent researches showed specific responses of the human's organism
to the exposure to solar UV irradiance that also could be taken into
account. For instance, some studies indicated that together with its
protective role, the melanin is able to enhance the probability for
triggering cancer processes through various biochemical pathways
induced by UV radiation \citep{Takeuchi2004,Premi2015}. Also, the
circadian rhythm together with the life style were found to impact
the erythema and DNA harms
\citep{Gaddameedhi2015,Sarkar_Gaddameedhi2018} that assumes an
adaptation of the permanent habitant's organism to the local
irradiative conditions. However, nowadays transport methods allow
very fast movement over the world, even within one day.
Long-distance moving across longitudes leads to an appreciable phase
shifting of the sleep-wake cycle with respect to the day-night
cycle. This occurrence could disturb the circadian rhythm and as a
result to impact the organism's response to the solar UV
irradiation. On the other hand, the movement across the latitudes
causes a sharp change in the solar UV radiation levels. In case of
traveling towards more irradiated geographical regions, such stress
could affect the usual intensity of bioprocesses including DNA
lesion-repair equilibrium. These conditions contribute to increasing
of damages and some of the alterations to the DNA may remain as
permanent mutations \citep{Roy_Gies2017}. To illustrate the solar
irradiance variations as a fuction of latitude, Fig. \ref{Fig4}
exhibits the diurnal time patterns of solar UV-B and UV-A radiations
evaluated by TUV model for clear sky conditions at three different
latitudes of the northern hemisphere. It can be seen that the
maximum of UV-B irradiance at local noon at $60^{\circ}$N is about
68\% lower than the maximum at $40^{\circ}$N and nearly twice lower
than at $20^{\circ}$N. The corresponding differences for UV-A
irradiance are smaller presenting $\sim 28$\% lower value at
$60^{\circ}$N with respect to $40^{\circ}$N and $\sim36$\% with
respect to $20^{\circ}$N. The increase of radiation maximum passing
from mid-latitudes ($40^{\circ}$) to tropics ($20^{\circ}$) is
$\sim17$\% for UV-B and $\sim6$\% for UV-A. Similar relations take
place for the UV-B and UV-A doses evaluated over a half-hour period
during the local noon. This example shows an appreciable increase in
UV-B and UV-A doses, when one moves from the polar regions to
mid-latitudes or tropics.

To enlarge the basis of protective strategy by including additional
effects like previously indicated, a proper parametrization of such
effects need to be performed.  A similar attempt was made by
\cite{Petkov2011} who evaluated the exposure times $\Delta
T_{UV-A,min}$ to solar UV-A irradiance that could lead to doses
$D_{UV-A,min}$, which were found to trigger SCC in laboratory
experiments with HaCaT keratinocytes
\citep{Colombo2017,Wischermann2008,He2006}. For this purpose,
weighting functions that replace the action spectrum in Eq.
(\ref{Eq1}) were constructed by using both spectrum of the UV-A
lamps applied in laboratory and the solar spectrum. Taking the doses
found to cause SCC in experiments as $D_{UV-A,min}$, these weighting
functions allowed the evaluation through the common approach
presented by Eqs. (\ref{Eq1}) and (\ref{Eq2}) of the times $\Delta
T_{UV-A,min}$ in real environmental conditions. Exposure periods
$\Delta T_{UV-A,min}$ up to a month were obtained adopting diverse
diurnal irradiation regimes and assuming a cumulative effect of
subexposures \citep{Wischermann2008,Lavker1995,Lai2018}. It should
be noted that these preliminary estimations just extrapolated the
results achieved in laboratory to real environmental conditions.
Their reliability that closely depends on the ability of HaCaT cells
to represent the DNA behaviour in the human organism needs
additional study. The example above illustrates a possible way for
parameterizing the results about bioprocesses in order to be used
for discussed enlargement of the protective strategy basis.

\section{Conclusions.}
\label{Sec5} The current protective strategy against the health
hazards associated with the exposure to solar UV radiation has been
constructed by using the erythema as a basic effect. Beside a brief
description of the strategy, the present study discussed whether the
messages about protection from the erythema had the same importance
for carcinogenesis triggered by solar radiation. It was assumed that
the risk of cancer formation decreases if the sunburning has been
avoided. However, such assumption was made on the basis of
epidemiological (statistical) studies without bearing in mind the
variety of pathways for DNA damages. Similar relationship between
erythema and carcinogenesis cannot provide any parametrization of
the second effect as in the case of the former. A parametrization
expressed by the action spectrum and minimal dose of weighted UV
radiation is needed to make a reliable estimation of the exposure
time for protective messages. Moreover, the different nature of the
erythema and the lesions in DNA caused by UV radiation strengthens
the doubt that above scheme could represent the probability of
cancer development. On the other hand, the weight of UV-A radiation
in erythema is quite low, while the role of UV-A for the cancer
formation is well evidenced in the latest studies. In addition, the
efficiency of the preventions would enhance if they take into
account also the particularities of the contemporaneous life, like
the possibility of fast transport from a place with low solar
irradiation to a place characterized by much higher irradiation.
These considerations suggest an enlargement of the protective
strategy basis by seeking a method to including also the DNA damages
beside erythema.

\section{Acknowledgements}
The help of Miss Luca Sainz Canipa is kindly acknowledged.


\begin{thebibliography}{86}
\expandafter\ifx\csname
natexlab\endcsname\relax\def\natexlab#1{#1}\fi
\providecommand{\url}[1]{\texttt{#1}} \providecommand{\href}[2]{#2}
\providecommand{\path}[1]{#1} \providecommand{\DOIprefix}{doi:}
\providecommand{\ArXivprefix}{arXiv:}
\providecommand{\URLprefix}{URL: }
\providecommand{\Pubmedprefix}{pmid:}
\providecommand{\doi}[1]{\href{http://dx.doi.org/#1}{\path{#1}}}
\providecommand{\Pubmed}[1]{\href{pmid:#1}{\path{#1}}}
\providecommand{\bibinfo}[2]{#2} \ifx\xfnm\relax
\def\xfnm[#1]{\unskip,\space#1}\fi
\bibitem[{Agache et~al.(1998)Agache, Quencez and Ota}]{Agache1998}
\bibinfo{author}{Agache, P.G.}, \bibinfo{author}{Quencez, E.},
  \bibinfo{author}{Ota, M.}, \bibinfo{year}{1998}.
\newblock \bibinfo{title}{The mechanism of solar erythema.}
\newblock \bibinfo{journal}{J Appl Cosmetol} \bibinfo{volume}{6},
  \bibinfo{pages}{69--78}.
\bibitem[{Agar et~al.(2004)Agar, Halliday, Barnetson, Ananthaswamy, Wheeler and
  AM}]{Agar2004}
\bibinfo{author}{Agar, N.S.}, \bibinfo{author}{Halliday, G.M.},
  \bibinfo{author}{Barnetson, R.S.}, \bibinfo{author}{Ananthaswamy, H.N.},
  \bibinfo{author}{Wheeler, M.}, \bibinfo{author}{AM, A.M.J.},
  \bibinfo{year}{2004}.
\newblock \bibinfo{title}{The basal layer in human squamous tumors harbors more
  {UVA} than {UVB} fingerprint mutations: a role for {UVA} in human skin
  carcinogenesis.}
\newblock \bibinfo{journal}{Proc Natl Acad Sci USA} \bibinfo{volume}{101},
  \bibinfo{pages}{4954--4959}.
\bibitem[{Ambach and Blumthaler(1993)}]{Ambach_Blumthaler1993}
\bibinfo{author}{Ambach, W.}, \bibinfo{author}{Blumthaler, M.},
  \bibinfo{year}{1993}.
\newblock \bibinfo{title}{Biological effectiveness of solar {UV} radiation in
  humans.}
\newblock \bibinfo{journal}{Cell and Mol Life Sci} \bibinfo{volume}{49},
  \bibinfo{pages}{747--753}.
\bibitem[{Anders et~al.(1995)Anders, Altheide, Kn\"{a}lmann and
  Tronnier}]{Anders1995}
\bibinfo{author}{Anders, A.}, \bibinfo{author}{Altheide, H.J.},
  \bibinfo{author}{Kn\"{a}lmann, M.}, \bibinfo{author}{Tronnier, H.},
  \bibinfo{year}{1995}.
\newblock \bibinfo{title}{Action spectrum for erythema in humans investigated
  with dye laser.}
\newblock \bibinfo{journal}{Photochem Photobiol} \bibinfo{volume}{61},
  \bibinfo{pages}{200--205}.
\bibitem[{Anderson and Parrish(1981)}]{Anderson_Parrish1981}
\bibinfo{author}{Anderson, R.R.}, \bibinfo{author}{Parrish, J.A.},
  \bibinfo{year}{1981}.
\newblock \bibinfo{title}{The optics of human skin.}
\newblock \bibinfo{journal}{The Journal of Investigative Dermatology}
  \bibinfo{volume}{77}, \bibinfo{pages}{13--19}.
\bibitem[{Antoniou et~al.(2008)Antoniou, Kosmadaki, Stratigos and
  Katsambas}]{Antoniou2008}
\bibinfo{author}{Antoniou, W.C.}, \bibinfo{author}{Kosmadaki, M.G.},
  \bibinfo{author}{Stratigos, A.}, \bibinfo{author}{Katsambas, A.D.},
  \bibinfo{year}{2008}.
\newblock \bibinfo{title}{Sunscreens-what's important to know.}
\newblock \bibinfo{journal}{JEADV} \bibinfo{volume}{22},
  \bibinfo{pages}{1110--1119}.
\bibitem[{Apalla et~al.(2017)Apalla, Lallas, Sotiriou, Lazaridou and
  Ioannides}]{Apalla2017}
\bibinfo{author}{Apalla, Z.}, \bibinfo{author}{Lallas, A.},
  \bibinfo{author}{Sotiriou, E.}, \bibinfo{author}{Lazaridou, E.},
  \bibinfo{author}{Ioannides, D.}, \bibinfo{year}{2017}.
\newblock \bibinfo{title}{Epidemiological trends in skin cancer.}
\newblock \bibinfo{journal}{Dermatol Pract Concept} \bibinfo{volume}{7},
  \bibinfo{pages}{1--6}.
\bibitem[{Bais et~al.(2015)Bais, McKenzie, Bernhard, Aucamp, Ilyas, Madronich
  and Tourpali}]{Bais2015}
\bibinfo{author}{Bais, A.F.}, \bibinfo{author}{McKenzie, R.L.},
  \bibinfo{author}{Bernhard, G.}, \bibinfo{author}{Aucamp, P.J.},
  \bibinfo{author}{Ilyas, M.}, \bibinfo{author}{Madronich, S.},
  \bibinfo{author}{Tourpali, K.}, \bibinfo{year}{2015}.
\newblock \bibinfo{title}{Ozone depletion and climate change: impacts on {UV}
  radiation.}
\newblock \bibinfo{journal}{Photochem Photobiol Sci} \bibinfo{volume}{14},
  \bibinfo{pages}{19--52}.
\bibitem[{Baker et~al.(2017)Baker, Marchetti, Karsili, Stavros and
  Ashfold}]{Baker2017}
\bibinfo{author}{Baker, L.A.}, \bibinfo{author}{Marchetti, B.},
  \bibinfo{author}{Karsili, T.N.V.}, \bibinfo{author}{Stavros, V.G.},
  \bibinfo{author}{Ashfold, M.N.R.}, \bibinfo{year}{2017}.
\newblock \bibinfo{title}{Photoprotection: extending lessons learned from
  studying natural sunscreens to the design of artificial sunscreen
  constituents.}
\newblock \bibinfo{journal}{Chemical Society Reviews} \bibinfo{volume}{46},
  \bibinfo{pages}{3770--3791}.
\bibitem[{Bernhard et~al.(2015)Bernhard, Arola, Dahlback, Fioletov,
  Heikkil\"{a}, Johnsen, Koskela, Lakkala, Svendby and Tamminen}]{Bernhard2015}
\bibinfo{author}{Bernhard, G.}, \bibinfo{author}{Arola, A.},
  \bibinfo{author}{Dahlback, A.}, \bibinfo{author}{Fioletov, V.},
  \bibinfo{author}{Heikkil\"{a}, A.}, \bibinfo{author}{Johnsen, B.},
  \bibinfo{author}{Koskela, T.}, \bibinfo{author}{Lakkala, K.},
  \bibinfo{author}{Svendby, T.}, \bibinfo{author}{Tamminen, J.},
  \bibinfo{year}{2015}.
\newblock \bibinfo{title}{Comparison of {OMI} {UV} observations with
  ground-based measurements at high northern latitudes.}
\newblock \bibinfo{journal}{Atmos Chem Phys} \bibinfo{volume}{15},
  \bibinfo{pages}{7391--7412}.
\bibitem[{Bieliauskiene et~al.(2019)Bieliauskiene, Holm-Schou, Philipsen,
  Murphy, Sboukis, Schwarz, Young and Wulf1}]{Bieliauskiene2019}
\bibinfo{author}{Bieliauskiene, G.}, \bibinfo{author}{Holm-Schou, A.S.S.},
  \bibinfo{author}{Philipsen, P.A.}, \bibinfo{author}{Murphy, G.M.},
  \bibinfo{author}{Sboukis, D.}, \bibinfo{author}{Schwarz, T.},
  \bibinfo{author}{Young, A.R.}, \bibinfo{author}{Wulf1, H.C.},
  \bibinfo{year}{2019}.
\newblock \bibinfo{title}{Measurements of sun sensitivity in five {E}uropean
  countries confirm the relative nature of {F}itzpatrick skin phototype scale.}
\newblock \bibinfo{journal}{Photodermatol Photoimmunol Photomed}
  \bibinfo{volume}{00}, \bibinfo{pages}{1--6}.
\bibitem[{Birch-Johansen et~al.(2010)Birch-Johansen, Jensen, Mortensen, Olesen
  and Kj{\ae}r}]{Birch-Johansen2010}
\bibinfo{author}{Birch-Johansen, F.}, \bibinfo{author}{Jensen, A.},
  \bibinfo{author}{Mortensen, L.}, \bibinfo{author}{Olesen, A.B.},
  \bibinfo{author}{Kj{\ae}r, S.K.}, \bibinfo{year}{2010}.
\newblock \bibinfo{title}{Trends in the incidence of nonmelanoma skin cancer in
  {D}enmark 1978-2007: Rapid incidence increase among young {D}anish women.}
\newblock \bibinfo{journal}{Atmos Chem Phys} \bibinfo{volume}{127},
  \bibinfo{pages}{2190--2198}.
\bibitem[{Boukamp(2005)}]{Boukamp2005}
\bibinfo{author}{Boukamp, P.}, \bibinfo{year}{2005}.
\newblock \bibinfo{title}{{UV}-induced skin cancer: similarities-variations.}
\newblock \bibinfo{journal}{J Dtsch Dermatol Ges} \bibinfo{volume}{3},
  \bibinfo{pages}{493--503}.
\bibitem[{Bouwstra et~al.(2003)Bouwstra, de~Graaff, Gooris, Nijsse, Wiechers
  and van Aelst}]{Bouwstra2003}
\bibinfo{author}{Bouwstra, J.A.}, \bibinfo{author}{de~Graaff, A.},
  \bibinfo{author}{Gooris, G.S.}, \bibinfo{author}{Nijsse, J.},
  \bibinfo{author}{Wiechers, J.}, \bibinfo{author}{van Aelst, A.C.},
  \bibinfo{year}{2003}.
\newblock \bibinfo{title}{Water distribution and related morphology in human
  stratum corneum at different hydration levels.}
\newblock \bibinfo{journal}{J Invest Dermatol} \bibinfo{volume}{120},
  \bibinfo{pages}{750--758}.
\bibitem[{Brasseur and Solomon(2005)}]{Brasseur_Solomon}
\bibinfo{author}{Brasseur, G.P.}, \bibinfo{author}{Solomon, S.},
  \bibinfo{year}{2005}.
\newblock \bibinfo{title}{Aeronomy of the middle atmosphere}.
\newblock \bibinfo{publisher}{Springer}.
\newblock \bibinfo{note}{151-264}.
\bibitem[{Brenner and Hearing(2008)}]{Brenner_Hearing2008}
\bibinfo{author}{Brenner, M.}, \bibinfo{author}{Hearing, V.J.},
  \bibinfo{year}{2008}.
\newblock \bibinfo{title}{The protective role of melanin against {UV} damage in
  human skin.}
\newblock \bibinfo{journal}{Photochem Photobiol} \bibinfo{volume}{84},
  \bibinfo{pages}{539--549}.
\bibitem[{Bruls et~al.(1984)Bruls, Slaper, Leun and Berrens}]{Bruls1984}
\bibinfo{author}{Bruls, W.A.G.}, \bibinfo{author}{Slaper, H.},
  \bibinfo{author}{Leun, J.C.V.D.}, \bibinfo{author}{Berrens, L.},
  \bibinfo{year}{1984}.
\newblock \bibinfo{title}{Transmission of human epidermis and stratum corneum
  as a function of thickness in the ultraviolet and visible wavelengths.}
\newblock \bibinfo{journal}{Photochem Photobiol} \bibinfo{volume}{40},
  \bibinfo{pages}{485--494}.
\bibitem[{Bustamante et~al.(2019)Bustamante, Hernandez-Ferrer, Tewari, Sarria,
  Harrison, Puigdecanet, Nonell, Kang, Friedl\"{a}nder, Estivill, Gonz\'{a}lez,
  Nieuwenhuijsen and Young}]{Bustamante2019}
\bibinfo{author}{Bustamante, M.}, \bibinfo{author}{Hernandez-Ferrer, C.},
  \bibinfo{author}{Tewari, A.}, \bibinfo{author}{Sarria, Y.},
  \bibinfo{author}{Harrison, G.I.}, \bibinfo{author}{Puigdecanet, E.},
  \bibinfo{author}{Nonell, L.}, \bibinfo{author}{Kang, W.},
  \bibinfo{author}{Friedl\"{a}nder, M.R.}, \bibinfo{author}{Estivill, X.},
  \bibinfo{author}{Gonz\'{a}lez, J.R.}, \bibinfo{author}{Nieuwenhuijsen, M.},
  \bibinfo{author}{Young, A.R.}, \bibinfo{year}{2019}.
\newblock \bibinfo{title}{Dose and time effects of solar-simulated ultraviolet
  radiation on the \textit{in vivo} human skin transcriptome.}
\newblock \bibinfo{journal}{British Journal of Dermatology}
  \bibinfo{volume}{DOI 10.1111/bjd.18527}.
\bibitem[{Callens et~al.(2016)Callens, van Eycken, Henau and
  Garmyn}]{Callens2016}
\bibinfo{author}{Callens, J.}, \bibinfo{author}{van Eycken, L.},
  \bibinfo{author}{Henau, K.}, \bibinfo{author}{Garmyn, M.},
  \bibinfo{year}{2016}.
\newblock \bibinfo{title}{Epidemiology of basal and squamous cell carcinoma in
  {B}elgium: The need for a uniform and compulsory registration.}
\newblock \bibinfo{journal}{J Eur Acad Dermatol Venereol} \bibinfo{volume}{30},
  \bibinfo{pages}{1912--1918}.
\bibitem[{Coblentz and Stair(1934)}]{Coblentz_Stair1934}
\bibinfo{author}{Coblentz, W.W.}, \bibinfo{author}{Stair, R.},
  \bibinfo{year}{1934}.
\newblock \bibinfo{title}{Data of the spectral erythemic reaction of the
  untanned human skin to ultraviolet radiation}.
\newblock \bibinfo{journal}{J Res} \bibinfo{volume}{12},
  \bibinfo{pages}{13--14}.
\bibitem[{Colombo et~al.(2017)Colombo, Sangiovanni, Maggio, Mattozzi, Zava,
  Corbett, Fumagalli, Carlino, Corsetto, Scaccabarozzi, Calvieri, Gismondi,
  Taramelli and Dell'Agli}]{Colombo2017}
\bibinfo{author}{Colombo, I.}, \bibinfo{author}{Sangiovanni, E.},
  \bibinfo{author}{Maggio, R.}, \bibinfo{author}{Mattozzi, C.},
  \bibinfo{author}{Zava, S.}, \bibinfo{author}{Corbett, Y.},
  \bibinfo{author}{Fumagalli, M.}, \bibinfo{author}{Carlino, C.},
  \bibinfo{author}{Corsetto, P.A.}, \bibinfo{author}{Scaccabarozzi, D.},
  \bibinfo{author}{Calvieri, S.}, \bibinfo{author}{Gismondi, A.},
  \bibinfo{author}{Taramelli, D.}, \bibinfo{author}{Dell'Agli, M.},
  \bibinfo{year}{2017}.
\newblock \bibinfo{title}{Ha{C}a{T} cells as a reliable in vitro
  differentiation model to dissect the inflammatory/repair response of human
  keratinocytes}.
\newblock \bibinfo{journal}{Mediators of Inflammation} \bibinfo{volume}{2017},
  \bibinfo{pages}{7435621}.
\bibitem[{Corona(1996)}]{Corona1996}
\bibinfo{author}{Corona, R.}, \bibinfo{year}{1996}.
\newblock \bibinfo{title}{Epidemiology of nonmelanoma skin cancers: a review.}
\newblock \bibinfo{journal}{Ann. Ist. Super. Sanit\`{a}} \bibinfo{volume}{32},
  \bibinfo{pages}{37--42}.
\bibitem[{Debacq-Chainiaux et~al.(2012)Debacq-Chainiaux, Leduc, Verbeke and
  Toussaint}]{Debacq-Chainiaux2012}
\bibinfo{author}{Debacq-Chainiaux, F.}, \bibinfo{author}{Leduc, C.},
  \bibinfo{author}{Verbeke, A.}, \bibinfo{author}{Toussaint, O.},
  \bibinfo{year}{2012}.
\newblock \bibinfo{title}{U{V}, stress and aging.}
\newblock \bibinfo{journal}{Dermato-Endocrinology} \bibinfo{volume}{4},
  \bibinfo{pages}{236--240}.
\bibitem[{Delinasios et~al.(2018)Delinasios, Karbaschi, Cooke and
  Young1}]{Delinasios2018}
\bibinfo{author}{Delinasios, G.J.}, \bibinfo{author}{Karbaschi, M.},
  \bibinfo{author}{Cooke, M.S.}, \bibinfo{author}{Young1, A.R.},
  \bibinfo{year}{2018}.
\newblock \bibinfo{title}{Vitamin {E} inhibits the {UVAI} induction of "light"
  and "dark" cyclobutane pyrimidine dimers, and oxidatively generated {DNA}
  damage, in keratinocytes.}
\newblock \bibinfo{journal}{Scientific Reports} \bibinfo{volume}{8},
  \bibinfo{pages}{423}.
\bibitem[{D'Mello et~al.(2016)D'Mello, Finlay, Baguley and
  Askarian-Amiri}]{D'Mello2016}
\bibinfo{author}{D'Mello, S.A.N.}, \bibinfo{author}{Finlay, G.J.},
  \bibinfo{author}{Baguley, B.C.}, \bibinfo{author}{Askarian-Amiri, M.E.},
  \bibinfo{year}{2016}.
\newblock \bibinfo{title}{Signaling pathways in melanogenesis.}
\newblock \bibinfo{journal}{Int J Mol Sci} \bibinfo{volume}{17},
  \bibinfo{pages}{1144 -- 1162}.
\bibitem[{Dornelles et~al.(2004)Dornelles, Goldim and Cestari}]{Dornelles2004}
\bibinfo{author}{Dornelles, S.}, \bibinfo{author}{Goldim, J.},
  \bibinfo{author}{Cestari, T.}, \bibinfo{year}{2004}.
\newblock \bibinfo{title}{Determination of the minimal erythema dose and
  colorimetric measurements as indicators of skin sensitivity to {UV-B}
  radiation.}
\newblock \bibinfo{journal}{Photochem Photobiol} \bibinfo{volume}{79},
  \bibinfo{pages}{540--544}.
\bibitem[{Egawa et~al.(2007)Egawa, Hirao and Takahashi}]{Egawa2007}
\bibinfo{author}{Egawa, M.}, \bibinfo{author}{Hirao, T.},
  \bibinfo{author}{Takahashi, M.}, \bibinfo{year}{2007}.
\newblock \bibinfo{title}{In vivo estimation of stratum corneum thickness from
  water concentration profiles obtained with {R}aman spectroscopy.}
\newblock \bibinfo{journal}{Acta Derm-Venereol} \bibinfo{volume}{87},
  \bibinfo{pages}{4--8}.
\bibitem[{Elwood and Jopson(1997)}]{Elwood_Jopson1997}
\bibinfo{author}{Elwood, J.M.}, \bibinfo{author}{Jopson, J.},
  \bibinfo{year}{1997}.
\newblock \bibinfo{title}{Melanoma and sun exposure: an overview of published
  studies.}
\newblock \bibinfo{journal}{Int J Cancer} \bibinfo{volume}{73},
  \bibinfo{pages}{198--203}.
\bibitem[{Fitzpatrick(1988)}]{Fitzpatrick1988}
\bibinfo{author}{Fitzpatrick, T.B.}, \bibinfo{year}{1988}.
\newblock \bibinfo{title}{The validity and practicality of sun-reactive skin
  types {I} through {VI}.}
\newblock \bibinfo{journal}{Arch. Dermatol.} \bibinfo{volume}{124},
  \bibinfo{pages}{869--871}.
\bibitem[{Flarer(1955)}]{Flarer1955}
\bibinfo{author}{Flarer, F.}, \bibinfo{year}{1955}.
\newblock \bibinfo{title}{The causes of inflammatory erythema.}
\newblock \bibinfo{journal}{J Invest Dermatol} \bibinfo{volume}{24},
  \bibinfo{pages}{201--209}.
\bibitem[{Gaddameedhi et~al.(2015)Gaddameedhi, Selby, Kemp, Ye and
  Sancar}]{Gaddameedhi2015}
\bibinfo{author}{Gaddameedhi, S.}, \bibinfo{author}{Selby, C.P.},
  \bibinfo{author}{Kemp, M.G.}, \bibinfo{author}{Ye, R.},
  \bibinfo{author}{Sancar, A.}, \bibinfo{year}{2015}.
\newblock \bibinfo{title}{The circadian clock controls sunburn apoptosis and
  erythema in mouse skin.}
\newblock \bibinfo{journal}{J Invest Dermatol} \bibinfo{volume}{135},
  \bibinfo{pages}{1119--1127}.
\bibitem[{Green et~al.(1985)Green, Siskind, Bain and Alexander}]{Green1985}
\bibinfo{author}{Green, A.}, \bibinfo{author}{Siskind, V.},
  \bibinfo{author}{Bain, C.}, \bibinfo{author}{Alexander, J.},
  \bibinfo{year}{1985}.
\newblock \bibinfo{title}{Sunburn and malignant melanoma.}
\newblock \bibinfo{journal}{Br J Cancer} \bibinfo{volume}{51},
  \bibinfo{pages}{393--397}.
\bibitem[{Gruijl et~al.(1993)Gruijl, Sterenborg, Forbes, Davies, Cole,
  Kelfkens, van Weelden, Slaper and van~der Leun}]{Gruijl1993}
\bibinfo{author}{Gruijl, F.R.}, \bibinfo{author}{Sterenborg, H.J.C.M.},
  \bibinfo{author}{Forbes, P.D.}, \bibinfo{author}{Davies, R.E.},
  \bibinfo{author}{Cole, C.}, \bibinfo{author}{Kelfkens, G.},
  \bibinfo{author}{van Weelden, H.}, \bibinfo{author}{Slaper, H.},
  \bibinfo{author}{van~der Leun, J.C.}, \bibinfo{year}{1993}.
\newblock \bibinfo{title}{Wavelength dependence of skin cancer induction by
  ultraviolet irradiation of albino hairless mice}.
\newblock \bibinfo{journal}{Cancer Res.} \bibinfo{volume}{53},
  \bibinfo{pages}{53--60}.
\bibitem[{Guy-Jr et~al.(2015)Guy-Jr, Thomas, Thompson, Watson, Massetti and
  Richardson}]{Guy2015}
\bibinfo{author}{Guy-Jr, G.P.}, \bibinfo{author}{Thomas, C.C.},
  \bibinfo{author}{Thompson, T.}, \bibinfo{author}{Watson, M.},
  \bibinfo{author}{Massetti, G.M.}, \bibinfo{author}{Richardson, L.C.},
  \bibinfo{year}{2015}.
\newblock \bibinfo{title}{Vital signs: Melanoma incidence and mortality trends
  and projections-{U}nited {S}tates, 1982-2030.}
\newblock \bibinfo{journal}{Morb Mortal Wkly Rep} \bibinfo{volume}{64},
  \bibinfo{pages}{591--596}.
\bibitem[{Harrison et~al.(2016)Harrison, Garz\'{o}n-Chavez and
  Nikles}]{Harrison2016}
\bibinfo{author}{Harrison, S.L.}, \bibinfo{author}{Garz\'{o}n-Chavez, D.R.},
  \bibinfo{author}{Nikles, C.J.}, \bibinfo{year}{2016}.
\newblock \bibinfo{title}{Sun protection policies of {A}ustralian primary
  schools in a region of high sun exposure.}
\newblock \bibinfo{journal}{Health Educ Res} \bibinfo{volume}{31},
  \bibinfo{pages}{416--428}.
\bibitem[{He et~al.(2006)He, Pi, Huang, Diwan, Waalkes and Chignel}]{He2006}
\bibinfo{author}{He, Y.Y.}, \bibinfo{author}{Pi, J.}, \bibinfo{author}{Huang,
  J.L.}, \bibinfo{author}{Diwan, B.A.}, \bibinfo{author}{Waalkes, M.P.},
  \bibinfo{author}{Chignel, C.F.}, \bibinfo{year}{2006}.
\newblock \bibinfo{title}{Chronic {UVA} irradiation of human {H}a{C}a{T}
  keratinocytes induces malignant transformation associated with acquired
  apoptotic resistance.}
\newblock \bibinfo{journal}{Oncogene} \bibinfo{volume}{25},
  \bibinfo{pages}{3680--3688}.
\bibitem[{Jhappan et~al.(2003)Jhappan, Noonan and Merlino}]{Jhappan2003}
\bibinfo{author}{Jhappan, C.}, \bibinfo{author}{Noonan, F.P.},
  \bibinfo{author}{Merlino, G.}, \bibinfo{year}{2003}.
\newblock \bibinfo{title}{Ultraviolet radiation and cutaneous malignant
  melanoma.}
\newblock \bibinfo{journal}{Oncogene} \bibinfo{volume}{22},
  \bibinfo{pages}{3099--3112}.
\bibitem[{Karran and Brem(2016)}]{Karran_Brem2016}
\bibinfo{author}{Karran, P.}, \bibinfo{author}{Brem, R.}, \bibinfo{year}{2016}.
\newblock \bibinfo{title}{Protein oxidation, {UVA} and human {DNA} repair.}
\newblock \bibinfo{journal}{DNA Repair} \bibinfo{volume}{44},
  \bibinfo{pages}{178--185}.
\bibitem[{Kaur and Saraf(2011)}]{Kaur_Saraf2011}
\bibinfo{author}{Kaur, C.D.}, \bibinfo{author}{Saraf, S.},
  \bibinfo{year}{2011}.
\newblock \bibinfo{title}{Skin care assessment on the basis of skin hydration,
  melanin, erythema and sebum at various body sites.}
\newblock \bibinfo{journal}{International Journal of Pharmacy and
  Pharmaceutical Sciences} \bibinfo{volume}{3}, \bibinfo{pages}{210--213}.
\bibitem[{Kiefer(2007)}]{Kiefer2007}
\bibinfo{author}{Kiefer, J.}, \bibinfo{year}{2007}.
\newblock \bibinfo{title}{Chromosomal Alterations: Methods, Results and
  Importance in Human Health}. \bibinfo{publisher}{Springer-Verlag Berlin
  Heidelberg}. chapter \bibinfo{chapter}{Effects of Ultraviolet Radiation on
  {DNA}}.
\newblock pp. \bibinfo{pages}{39--53}.
\bibitem[{Lai et~al.(2018)Lai, Cranwell and Sinclair}]{Lai2018}
\bibinfo{author}{Lai, V.}, \bibinfo{author}{Cranwell, W.},
  \bibinfo{author}{Sinclair, R.}, \bibinfo{year}{2018}.
\newblock \bibinfo{title}{Epidemiology of skin cancer in the mature patient.}
\newblock \bibinfo{journal}{Clinics in Dermatology} \bibinfo{volume}{36},
  \bibinfo{pages}{167--176}.
\bibitem[{Lavker et~al.(1995)Lavker, Gerberick, Veres, Irwin and
  Kaidbey}]{Lavker1995}
\bibinfo{author}{Lavker, R.M.}, \bibinfo{author}{Gerberick, G.F.},
  \bibinfo{author}{Veres, D.}, \bibinfo{author}{Irwin, C.J.},
  \bibinfo{author}{Kaidbey, K.H.}, \bibinfo{year}{1995}.
\newblock \bibinfo{title}{Cumulative effects from repeated exposures to
  suberythemal doses of {UVB} and {UVA} in human skin.}
\newblock \bibinfo{journal}{Journal of the American Academy of Dermatology}
  \bibinfo{volume}{32}, \bibinfo{pages}{53--62}.
\bibitem[{Lehmann et~al.(2011)Lehmann, McGibbon and Stefanini}]{Lehmann2011}
\bibinfo{author}{Lehmann, A.R.}, \bibinfo{author}{McGibbon, D.},
  \bibinfo{author}{Stefanini, M.}, \bibinfo{year}{2011}.
\newblock \bibinfo{title}{Xeroderma pigmentosum.}
\newblock \bibinfo{journal}{Orphanet Journal of Rare Diseases}
  \bibinfo{volume}{6}, \bibinfo{pages}{70}.
\bibitem[{Lucas et~al.(2018)Lucas, Neale, Madronich and McKenzie}]{Lucas2018}
\bibinfo{author}{Lucas, R.M.}, \bibinfo{author}{Neale, R.E.},
  \bibinfo{author}{Madronich, S.}, \bibinfo{author}{McKenzie, R.L.},
  \bibinfo{year}{2018}.
\newblock \bibinfo{title}{Are current guidelines for sun protection optimal for
  health? {E}xploring the evidence.}
\newblock \bibinfo{journal}{Photochem Photobiol Sci} \bibinfo{volume}{17},
  \bibinfo{pages}{1956 -- 1963}.
\bibitem[{Luckiesh et~al.(1930)Luckiesh, Holladay and Taylor}]{Luckiesh1930}
\bibinfo{author}{Luckiesh, M.}, \bibinfo{author}{Holladay, L.L.},
  \bibinfo{author}{Taylor, A.H.}, \bibinfo{year}{1930}.
\newblock \bibinfo{title}{Reaction of untanned human skin to ultraviolet
  radiation.}
\newblock \bibinfo{journal}{J Opt Soc Am} \bibinfo{volume}{20},
  \bibinfo{pages}{423--432}.
\bibitem[{MacLaughlin et~al.(1982)MacLaughlin, Anderson and
  Holick}]{MacLaughlin1982}
\bibinfo{author}{MacLaughlin, J.A.}, \bibinfo{author}{Anderson, R.R.},
  \bibinfo{author}{Holick, M.F.}, \bibinfo{year}{1982}.
\newblock \bibinfo{title}{Spectral character of sunlight modulates
  photosynthesis of previtamin {D}3 and its photoisometers in human skin}.
\newblock \bibinfo{journal}{Science} \bibinfo{volume}{216},
  \bibinfo{pages}{1001--1003}.
\bibitem[{Maddodi et~al.(2012)Maddodi, Jayanthy and Setaluri}]{Maddodi2012}
\bibinfo{author}{Maddodi, N.}, \bibinfo{author}{Jayanthy, A.},
  \bibinfo{author}{Setaluri, V.}, \bibinfo{year}{2012}.
\newblock \bibinfo{title}{Shining light on skin pigmentation: The darker and
  the brighter side of effects of {UV} radiation.}
\newblock \bibinfo{journal}{Photochem Photobiol} \bibinfo{volume}{88},
  \bibinfo{pages}{1075--1082}.
\bibitem[{Madronich and Flocke(1997)}]{Madronich_Flocke1997}
\bibinfo{author}{Madronich, S.}, \bibinfo{author}{Flocke, S.},
  \bibinfo{year}{1997}.
\newblock \bibinfo{title}{Solar ultraviolet radiation - modeling, measurements
  and effects}. \bibinfo{publisher}{NATO ASI series,Springer, Berlin}. volume
  \bibinfo{volume}{I52}.
\bibitem[{McAdam et~al.(2016)McAdam, Brem and Karran}]{McAdam2016}
\bibinfo{author}{McAdam, E.}, \bibinfo{author}{Brem, R.},
  \bibinfo{author}{Karran, P.}, \bibinfo{year}{2016}.
\newblock \bibinfo{title}{Oxidative stress-induced protein damage inhibits
  {DNA} repair and determines mutation risk and therapeutic efficacy.}
\newblock \bibinfo{journal}{Mol Cancer Res} \bibinfo{volume}{14},
  \bibinfo{pages}{612--622}.
\bibitem[{McKenzie et~al.(2009)McKenzie, Liley and Bj\"{o}rn}]{McKenzie2009}
\bibinfo{author}{McKenzie, R.L.}, \bibinfo{author}{Liley, J.B.},
  \bibinfo{author}{Bj\"{o}rn, L.O.}, \bibinfo{year}{2009}.
\newblock \bibinfo{title}{U{V} radiation: balancing risks and benefits.}
\newblock \bibinfo{journal}{Photochem Photobiol} \bibinfo{volume}{85},
  \bibinfo{pages}{88--98}.
\bibitem[{McKinlay and Diffey(1987)}]{McKinlay_Diffey1987}
\bibinfo{author}{McKinlay, A.F.}, \bibinfo{author}{Diffey, B.L.},
  \bibinfo{year}{1987}.
\newblock \bibinfo{title}{A reference action spectrum for ultraviolet induced
  erythema in human skin.}
\newblock \bibinfo{journal}{CIE J} \bibinfo{volume}{6},
  \bibinfo{pages}{17--22}.
\bibitem[{Melnikova and Vasilyev(2005)}]{Melnikova_Vasilyev}
\bibinfo{author}{Melnikova, I.N.}, \bibinfo{author}{Vasilyev, A.V.},
  \bibinfo{year}{2005}.
\newblock \bibinfo{title}{Short-Wave Solar Radiation in the Earth's
  Atmosphere}.
\newblock \bibinfo{publisher}{Springer-Verlag Berlin Heidelberg}.
\bibitem[{Mims-III et~al.(2019)Mims-III, McGonigle, Wilkes, Parisi, Grant, Cook
  and Pering}]{Mims2019}
\bibinfo{author}{Mims-III, F.M.}, \bibinfo{author}{McGonigle, A.J.S.},
  \bibinfo{author}{Wilkes, T.C.}, \bibinfo{author}{Parisi, A.V.},
  \bibinfo{author}{Grant, W.B.}, \bibinfo{author}{Cook, J.M.},
  \bibinfo{author}{Pering, T.D.}, \bibinfo{year}{2019}.
\newblock \bibinfo{title}{Measuring and visualizing solar uv for a wide range
  of atmospheric conditions on {H}awai'i island.}
\newblock \bibinfo{journal}{Int J Environ Res Public Health}
  \bibinfo{volume}{16}, \bibinfo{pages}{997}.
\bibitem[{Mouret et~al.(2008)Mouret, Charveron, Favier, Cadet and
  Douki}]{Mouret2008}
\bibinfo{author}{Mouret, S.}, \bibinfo{author}{Charveron, M.},
  \bibinfo{author}{Favier, A.}, \bibinfo{author}{Cadet, J.},
  \bibinfo{author}{Douki, T.}, \bibinfo{year}{2008}.
\newblock \bibinfo{title}{Differential repair of {UVB}-induced cyclobutane
  pyrimidine dimers in cultured human skin cells and whole human skin.}
\newblock \bibinfo{journal}{DNA Repair} \bibinfo{volume}{7},
  \bibinfo{pages}{704--712}.
\bibitem[{Mullenders(2018)}]{Mullenders2018}
\bibinfo{author}{Mullenders, L.H.F.}, \bibinfo{year}{2018}.
\newblock \bibinfo{title}{Solar {UV} damage to cellular {DNA}: from mechanisms
  to biological effects.}
\newblock \bibinfo{journal}{Photochem Photobiol Sci} \bibinfo{volume}{17},
  \bibinfo{pages}{1842--1852}.
\bibitem[{Nishigori(2014)}]{Nishigori2014}
\bibinfo{author}{Nishigori, C.}, \bibinfo{year}{2014}.
\newblock \bibinfo{title}{Cancer and inflammation mechanisms.}.
  \bibinfo{publisher}{John Wiley \& Sons Inc., Hoboken, New Jersey}. chapter
  \bibinfo{chapter}{Photocarcinogenesis and inflammation}.
\newblock pp. \bibinfo{pages}{271--283}.
\bibitem[{Passeron et~al.(2019)Passeron, Bouillon, Callender, Cestari, Diepgen,
  Green, van~der Pols, Bernard, Ly, Bernerd, Marrot, Nielsen, Verschoore,
  Jablonski and Young}]{Passeron2019}
\bibinfo{author}{Passeron, T.}, \bibinfo{author}{Bouillon, R.},
  \bibinfo{author}{Callender, V.}, \bibinfo{author}{Cestari, T.},
  \bibinfo{author}{Diepgen, T.L.}, \bibinfo{author}{Green, A.C.},
  \bibinfo{author}{van~der Pols, J.C.}, \bibinfo{author}{Bernard, B.A.},
  \bibinfo{author}{Ly, F.}, \bibinfo{author}{Bernerd, F.},
  \bibinfo{author}{Marrot, L.}, \bibinfo{author}{Nielsen, M.},
  \bibinfo{author}{Verschoore, M.}, \bibinfo{author}{Jablonski, N.G.},
  \bibinfo{author}{Young, A.R.}, \bibinfo{year}{2019}.
\newblock \bibinfo{title}{Sunscreen photoprotection and vitamin {D} status.}
\newblock \bibinfo{journal}{British Journal of Dermatology}
  \bibinfo{volume}{181}, \bibinfo{pages}{916--931}.
\bibitem[{Petkov et~al.(2011)Petkov, Vitale, Tomasi, Gadaleta, Mazzola,
  Lanconelli, Lupi, Busetto and Benedetti}]{Petkov2011}
\bibinfo{author}{Petkov, B.}, \bibinfo{author}{Vitale, V.},
  \bibinfo{author}{Tomasi, C.}, \bibinfo{author}{Gadaleta, E.},
  \bibinfo{author}{Mazzola, M.}, \bibinfo{author}{Lanconelli, C.},
  \bibinfo{author}{Lupi, A.}, \bibinfo{author}{Busetto, M.},
  \bibinfo{author}{Benedetti, E.}, \bibinfo{year}{2011}.
\newblock \bibinfo{title}{Preliminary assessment of the risks associated with
  solar ultraviolet-{A} exposure.}
\newblock \bibinfo{journal}{Rad Environ Biophys} \bibinfo{volume}{50},
  \bibinfo{pages}{219--229}.
\bibitem[{Petkov et~al.(2014a)Petkov, Vitale, Tomasi, Siani, Seckmeyer, Webb,
  Smedley, Casale, Werner, Lanconelli, Mazzola, Lupi, Busetto, Di\'{e}moz,
  Goutail, K\"{o}hler, Mendeva, Josefsson, Moore, Bartolom\'{e}, Gonz\'{a}lez,
  Mi\u{s}aga, Dahlback, T\'{o}th, Varghese, de~Backer, St\"{u}bi and
  Van\'{\i}\u{c}ek}]{Petkov2014}
\bibinfo{author}{Petkov, B.H.}, \bibinfo{author}{Vitale, V.},
  \bibinfo{author}{Tomasi, C.}, \bibinfo{author}{Siani, A.M.},
  \bibinfo{author}{Seckmeyer, G.}, \bibinfo{author}{Webb, A.R.},
  \bibinfo{author}{Smedley, A.R.D.}, \bibinfo{author}{Casale, G.R.},
  \bibinfo{author}{Werner, R.}, \bibinfo{author}{Lanconelli, C.},
  \bibinfo{author}{Mazzola, M.}, \bibinfo{author}{Lupi, A.},
  \bibinfo{author}{Busetto, M.}, \bibinfo{author}{Di\'{e}moz, H.},
  \bibinfo{author}{Goutail, F.}, \bibinfo{author}{K\"{o}hler, U.},
  \bibinfo{author}{Mendeva, B.D.}, \bibinfo{author}{Josefsson, W.},
  \bibinfo{author}{Moore, D.}, \bibinfo{author}{Bartolom\'{e}, M.L.},
  \bibinfo{author}{Gonz\'{a}lez, J.R.M.}, \bibinfo{author}{Mi\u{s}aga, O.},
  \bibinfo{author}{Dahlback, A.}, \bibinfo{author}{T\'{o}th, Z.},
  \bibinfo{author}{Varghese, S.}, \bibinfo{author}{de~Backer, H.},
  \bibinfo{author}{St\"{u}bi, R.}, \bibinfo{author}{Van\'{\i}\u{c}ek, K.},
  \bibinfo{year}{2014a}.
\newblock \bibinfo{title}{Response of the ozone column over {E}urope to the
  2011 {A}rctic ozone depletion event according to groundbased observations and
  assessment of the consequent variations in surface {UV} irradiance.}
\newblock \bibinfo{journal}{Atmos Environ} \bibinfo{volume}{85},
  \bibinfo{pages}{169--178}.
\bibitem[{Premi et~al.(2015)Premi, Wallisch, Mano, Weiner, Bacchiocchi,
  Wakamatsu, Bechara, Halaban, Douki and Bras}]{Premi2015}
\bibinfo{author}{Premi, S.}, \bibinfo{author}{Wallisch, S.},
  \bibinfo{author}{Mano, C.M.}, \bibinfo{author}{Weiner, A.B.},
  \bibinfo{author}{Bacchiocchi, A.}, \bibinfo{author}{Wakamatsu, K.},
  \bibinfo{author}{Bechara, E.J.H.}, \bibinfo{author}{Halaban, R.},
  \bibinfo{author}{Douki, T.}, \bibinfo{author}{Bras, D.E.},
  \bibinfo{year}{2015}.
\newblock \bibinfo{title}{Chemiexcitation of melanin derivatives induces {DNA}
  photoproducts long after {UV} exposure.}
\newblock \bibinfo{journal}{Science} \bibinfo{volume}{347}, \bibinfo{pages}{842
  -- 847}.
\bibitem[{Rastogi et~al.(2010)Rastogi, Richa, Kumar, Tyagi and
  Sinha}]{Rastogi2010}
\bibinfo{author}{Rastogi, R.P.}, \bibinfo{author}{Richa},
  \bibinfo{author}{Kumar, A.}, \bibinfo{author}{Tyagi, M.B.},
  \bibinfo{author}{Sinha, R.P.}, \bibinfo{year}{2010}.
\newblock \bibinfo{title}{Molecular mechanisms of ultraviolet radiation-induced
  {DNA} damage and repair.}
\newblock \bibinfo{journal}{Journal of Nucleic Acids} \bibinfo{volume}{article
  ID 592980}, \bibinfo{pages}{32}.
\bibitem[{Ravanat et~al.(2001)Ravanat, Douki and Cadet}]{Ravanat2001}
\bibinfo{author}{Ravanat, J.L.}, \bibinfo{author}{Douki, T.},
  \bibinfo{author}{Cadet, J.}, \bibinfo{year}{2001}.
\newblock \bibinfo{title}{Direct and indirect effects of {UV} radiation on
  {DNA} and its components.}
\newblock \bibinfo{journal}{J Photochem Photobiol B} \bibinfo{volume}{63},
  \bibinfo{pages}{88--102}.
\bibitem[{Reichrath(2014)}]{Reichrath2014}
\bibinfo{editor}{Reichrath, J.} (Ed.), \bibinfo{year}{2014}.
\newblock \bibinfo{title}{Sunlight, vitamin {D} and skin cancer}.
\newblock \bibinfo{publisher}{Springer, New York, USA}.
\bibitem[{Roy and Gies(2017)}]{Roy_Gies2017}
\bibinfo{author}{Roy, C.}, \bibinfo{author}{Gies, P.}, \bibinfo{year}{2017}.
\newblock \bibinfo{title}{Non-ionizing Radiation Protection.}.
  \bibinfo{publisher}{John Wiley \& Sons, Inc., Hoboken, USA}. chapter
  \bibinfo{chapter}{U{VR} and Short-Term Hazards to the Skin and Eyes.}
\newblock pp. \bibinfo{pages}{49 -- 66}.
\bibitem[{Sarkar and Gaddameedhi(2018)}]{Sarkar_Gaddameedhi2018}
\bibinfo{author}{Sarkar, S.}, \bibinfo{author}{Gaddameedhi, S.},
  \bibinfo{year}{2018}.
\newblock \bibinfo{title}{U{V}-{B}-induced erythema in human skin: The
  circadian clock is ticking.}
\newblock \bibinfo{journal}{Journal of Investigative Dermatology}
  \bibinfo{volume}{138}, \bibinfo{pages}{248 -- 251}.
\bibitem[{Schmalwieser et~al.(2017)Schmalwieser, Gr\"{o}bner, Blumthaler,
  Klotz, de~Backer, Bols\'{e}e, Werner, Tomsic, Metelka, Eriksen, Jepsen, Aun,
  Heikkil\"{a}, Duprat, Sandmann, Weiss, Bais, Toth, Siani, Vaccaro,
  Di\'{e}moz, Grifoni, Zipoli, Lorenzetto, Petkov, di~Sarra, Massen, Yousif,
  Aculinin, den Outer, Svendby, Dahlback, Johnsen, Biszczuk-Jakubowska,
  Krzyscin, Henriques, Chubarova, Kolar\u{z}, Mijatovic, Groselj, Pribullova,
  Gonzales, Bilbao, Guerrero, Serrano, Andersson, Vuilleumier, Webb and
  O'Hagan}]{Schmalwieser2017}
\bibinfo{author}{Schmalwieser, A.W.}, \bibinfo{author}{Gr\"{o}bner, J.},
  \bibinfo{author}{Blumthaler, M.}, \bibinfo{author}{Klotz, B.},
  \bibinfo{author}{de~Backer, H.}, \bibinfo{author}{Bols\'{e}e, D.},
  \bibinfo{author}{Werner, R.}, \bibinfo{author}{Tomsic, D.},
  \bibinfo{author}{Metelka, L.}, \bibinfo{author}{Eriksen, P.},
  \bibinfo{author}{Jepsen, N.}, \bibinfo{author}{Aun, M.},
  \bibinfo{author}{Heikkil\"{a}, A.}, \bibinfo{author}{Duprat, T.},
  \bibinfo{author}{Sandmann, H.}, \bibinfo{author}{Weiss, T.},
  \bibinfo{author}{Bais, A.}, \bibinfo{author}{Toth, Z.},
  \bibinfo{author}{Siani, A.M.}, \bibinfo{author}{Vaccaro, L.},
  \bibinfo{author}{Di\'{e}moz, H.}, \bibinfo{author}{Grifoni, D.},
  \bibinfo{author}{Zipoli, G.}, \bibinfo{author}{Lorenzetto, G.},
  \bibinfo{author}{Petkov, B.H.}, \bibinfo{author}{di~Sarra, G.A.},
  \bibinfo{author}{Massen, F.}, \bibinfo{author}{Yousif, C.},
  \bibinfo{author}{Aculinin, A.A.}, \bibinfo{author}{den Outer, P.},
  \bibinfo{author}{Svendby, T.}, \bibinfo{author}{Dahlback, A.},
  \bibinfo{author}{Johnsen, B.}, \bibinfo{author}{Biszczuk-Jakubowska, J.},
  \bibinfo{author}{Krzyscin, J.}, \bibinfo{author}{Henriques, D.},
  \bibinfo{author}{Chubarova, N.}, \bibinfo{author}{Kolar\u{z}, P.},
  \bibinfo{author}{Mijatovic, Z.}, \bibinfo{author}{Groselj, D.},
  \bibinfo{author}{Pribullova, A.}, \bibinfo{author}{Gonzales, J.R.M.},
  \bibinfo{author}{Bilbao, J.}, \bibinfo{author}{Guerrero, J.M.V.},
  \bibinfo{author}{Serrano, A.}, \bibinfo{author}{Andersson, S.},
  \bibinfo{author}{Vuilleumier, L.}, \bibinfo{author}{Webb, A.},
  \bibinfo{author}{O'Hagan, J.}, \bibinfo{year}{2017}.
\newblock \bibinfo{title}{U{V} {I}ndex monitoring in {E}urope.}
\newblock \bibinfo{journal}{Photochem Photobio Sci} \bibinfo{volume}{16},
  \bibinfo{pages}{1349--1370}.
\bibitem[{Seckmeyer et~al.(2012)Seckmeyer, Klingebiel, Riechelmann, Lohse,
  McKenzie, Liley, Allen, Siani and Casale}]{Seckmeyer2012}
\bibinfo{author}{Seckmeyer, G.}, \bibinfo{author}{Klingebiel, M.},
  \bibinfo{author}{Riechelmann, S.}, \bibinfo{author}{Lohse, I.},
  \bibinfo{author}{McKenzie, R.L.}, \bibinfo{author}{Liley, J.B.},
  \bibinfo{author}{Allen, M.W.}, \bibinfo{author}{Siani, A.M.},
  \bibinfo{author}{Casale, G.R.}, \bibinfo{year}{2012}.
\newblock \bibinfo{title}{A critical assessment of two types of personal {UV}
  dosimeters.}
\newblock \bibinfo{journal}{Photochem Photobiol} \bibinfo{volume}{88},
  \bibinfo{pages}{215--222}.
\bibitem[{Seckmeyer et~al.(2013)Seckmeyer, Schrempf, Wieczorek, Riechelmann,
  Graw, Seckmeyer and Zankl}]{Seckmeyer2013}
\bibinfo{author}{Seckmeyer, G.}, \bibinfo{author}{Schrempf, M.},
  \bibinfo{author}{Wieczorek, A.}, \bibinfo{author}{Riechelmann, S.},
  \bibinfo{author}{Graw, K.}, \bibinfo{author}{Seckmeyer, S.},
  \bibinfo{author}{Zankl, M.}, \bibinfo{year}{2013}.
\newblock \bibinfo{title}{A novel method to calculate solar {UV} exposure
  relevant to vitamin {D} production in humans.}
\newblock \bibinfo{journal}{Photochem Photobiol} \bibinfo{volume}{89},
  \bibinfo{pages}{974--983}.
\bibitem[{Setlow et~al.(1993)Setlow, Grist, Thompson and Woodhead}]{Setlow1993}
\bibinfo{author}{Setlow, R.}, \bibinfo{author}{Grist, E.},
  \bibinfo{author}{Thompson, K.}, \bibinfo{author}{Woodhead, A.D.},
  \bibinfo{year}{1993}.
\newblock \bibinfo{title}{Wavelengths effective in induction of malignant
  melanoma.}
\newblock \bibinfo{journal}{Proc Natl Acad Sci USA} \bibinfo{volume}{90},
  \bibinfo{pages}{6666--6670}.
\bibitem[{Setlow(1974)}]{Setlow1974}
\bibinfo{author}{Setlow, R.B.}, \bibinfo{year}{1974}.
\newblock \bibinfo{title}{The wavelengths in sunlight effective in producing
  skin cancer: a theoretical analysis.}
\newblock \bibinfo{journal}{Proc Natl Acad Sci USA} \bibinfo{volume}{71},
  \bibinfo{pages}{3363--3366}.
\bibitem[{Siani et~al.(2008)Siani, Casale, Di\'{e}moz, Agnesod, Kimlin, Lang
  and Colosimo}]{Siani2008}
\bibinfo{author}{Siani, A.M.}, \bibinfo{author}{Casale, G.R.},
  \bibinfo{author}{Di\'{e}moz, H.}, \bibinfo{author}{Agnesod, G.},
  \bibinfo{author}{Kimlin, M.G.}, \bibinfo{author}{Lang, C.A.},
  \bibinfo{author}{Colosimo, A.}, \bibinfo{year}{2008}.
\newblock \bibinfo{title}{Personal {UV} exposure in high albedo alpine sites.}
\newblock \bibinfo{journal}{Atmos Chem Phys} \bibinfo{volume}{8},
  \bibinfo{pages}{3749--3760}.
\bibitem[{Sinha and H\"{a}der(2002)}]{Sinha2002}
\bibinfo{author}{Sinha, R.P.}, \bibinfo{author}{H\"{a}der, D.P.},
  \bibinfo{year}{2002}.
\newblock \bibinfo{title}{U{V}-induced {DNA} damage and repair: a review}.
\newblock \bibinfo{journal}{Photochem Photobiol Sci} \bibinfo{volume}{1},
  \bibinfo{pages}{225--236}.
\bibitem[{Smith et~al.(2000)Smith, Ford, Hollander, Bortnick, Amundson, Seo,
  Deng, Hanawalt and jr}]{Smith2000}
\bibinfo{author}{Smith, M.L.}, \bibinfo{author}{Ford, J.M.},
  \bibinfo{author}{Hollander, M.C.}, \bibinfo{author}{Bortnick, R.A.},
  \bibinfo{author}{Amundson, S.A.}, \bibinfo{author}{Seo, Y.R.},
  \bibinfo{author}{Deng, C.X.}, \bibinfo{author}{Hanawalt, P.C.},
  \bibinfo{author}{jr, A.J.F.}, \bibinfo{year}{2000}.
\newblock \bibinfo{title}{p53-mediated {DNA} repair responses to {UV}
  radiation: studies of mouse cells lacking p53, p21, and/or gadd45 genes.}
\newblock \bibinfo{journal}{Mol Cell Biol} \bibinfo{volume}{20},
  \bibinfo{pages}{3705 -- 3714}.
\bibitem[{Solano(2014)}]{Solano2014}
\bibinfo{author}{Solano, F.}, \bibinfo{year}{2014}.
\newblock \bibinfo{title}{Melanins: Skin pigments and much more-types,
  structural models, biological functions, and formation routes.}
\newblock \bibinfo{journal}{New Journal of Science} \bibinfo{volume}{Article ID
  498276}, \bibinfo{pages}{28}.
\bibitem[{Tadokoro et~al.(2005)Tadokoro, Yamaguchi, Batzer, Coelho, Zmudzka,
  Miller, Wolber, Beer and Hearing}]{Tadokoro2005}
\bibinfo{author}{Tadokoro, T.}, \bibinfo{author}{Yamaguchi, Y.},
  \bibinfo{author}{Batzer, J.}, \bibinfo{author}{Coelho, S.G.},
  \bibinfo{author}{Zmudzka, B.Z.}, \bibinfo{author}{Miller, S.A.},
  \bibinfo{author}{Wolber, R.}, \bibinfo{author}{Beer, J.Z.},
  \bibinfo{author}{Hearing, V.J.}, \bibinfo{year}{2005}.
\newblock \bibinfo{title}{Mechanisms of skin tanning in different racial/ethnic
  groups in response to ultraviolet radiation.}
\newblock \bibinfo{journal}{J Invest Dermatol} \bibinfo{volume}{124},
  \bibinfo{pages}{1326--1332}.
\bibitem[{Takeuchi et~al.(2004)Takeuchi, Zhang, Wakamatsu, Ito, Hearing,
  Kraemer and Brash}]{Takeuchi2004}
\bibinfo{author}{Takeuchi, S.}, \bibinfo{author}{Zhang, W.},
  \bibinfo{author}{Wakamatsu, K.}, \bibinfo{author}{Ito, S.},
  \bibinfo{author}{Hearing, V.J.}, \bibinfo{author}{Kraemer, K.H.},
  \bibinfo{author}{Brash, D.E.}, \bibinfo{year}{2004}.
\newblock \bibinfo{title}{Melanin acts as a potent {UVB} photosensitizer to
  cause an atypical mode of cell death in murine skin.}
\newblock \bibinfo{journal}{PNAS} \bibinfo{volume}{101}, \bibinfo{pages}{15076
  -- 15081}.
\bibitem[{Tanskanen et~al.(2006)Tanskanen, Krotkov, Herman and
  Arola}]{Tanskanen2006}
\bibinfo{author}{Tanskanen, A.}, \bibinfo{author}{Krotkov, N.A.},
  \bibinfo{author}{Herman, J.R.}, \bibinfo{author}{Arola, A.},
  \bibinfo{year}{2006}.
\newblock \bibinfo{title}{Surface ultraviolet irradiance from {OMI}.}
\newblock \bibinfo{journal}{IEEE Transactions on Geoscience and Remote Sensing}
  \bibinfo{volume}{44}, \bibinfo{pages}{1267--1271}.
\bibitem[{Tucker(2008)}]{Tucker2008}
\bibinfo{author}{Tucker, M.A.}, \bibinfo{year}{2008}.
\newblock \bibinfo{title}{Is sunlight important to melanoma causation?}
\newblock \bibinfo{journal}{Cancer Epidemiol Biomarkers Prev}
  \bibinfo{volume}{17}, \bibinfo{pages}{467--468}.
\bibitem[{Wadhera et~al.(2006)Wadhera, Fazio, Bricca and Stanton}]{Wadhera2006}
\bibinfo{author}{Wadhera, A.}, \bibinfo{author}{Fazio, M.},
  \bibinfo{author}{Bricca, G.}, \bibinfo{author}{Stanton, O.},
  \bibinfo{year}{2006}.
\newblock \bibinfo{title}{Metastatic basal cell carcinoma: A case report and
  literature review. {H}ow accurate is our incidence data?}
\newblock \bibinfo{journal}{Dermatol Online J} \bibinfo{volume}{12},
  \bibinfo{pages}{7}.
\bibitem[{Wallingford et~al.(2015)Wallingford, Iannacone, Youlden, Baade, Ives,
  Verne, Aitken and Green}]{Wallingford2015}
\bibinfo{author}{Wallingford, S.C.}, \bibinfo{author}{Iannacone, M.R.},
  \bibinfo{author}{Youlden, D.R.}, \bibinfo{author}{Baade, P.D.},
  \bibinfo{author}{Ives, A.}, \bibinfo{author}{Verne, J.},
  \bibinfo{author}{Aitken, J.F.}, \bibinfo{author}{Green, A.C.},
  \bibinfo{year}{2015}.
\newblock \bibinfo{title}{Comparison of melanoma incidence and trends among
  youth under 25 years in {A}ustralia and {E}ngland, 1990-2010.}
\newblock \bibinfo{journal}{Int J Cancer} \bibinfo{volume}{137},
  \bibinfo{pages}{2227--2233}.
\bibitem[{WHO(2002)}]{WHO2002}
\bibinfo{author}{WHO}, \bibinfo{year}{2002}.
\newblock \bibinfo{title}{Global solar {UV} {I}ndex: A practical guide}.
\newblock \bibinfo{journal}{World Health Organization} .
\bibitem[{WHO(2003)}]{WHO2003}
\bibinfo{author}{WHO}, \bibinfo{year}{2003}.
\newblock \bibinfo{title}{I{NTERSUN} the global {UV} project: a guide and
  compendium.}
\newblock \bibinfo{journal}{World Health Organization} .
\bibitem[{Wischermann et~al.(2008)Wischermann, Popp, Moshir,
  Scharfetter-Kochanek, Wlaschek, de~Gruijl, Hartschuh, Greinert, Volkmer,
  Faust, Rapp, Schmezer and Boukamp}]{Wischermann2008}
\bibinfo{author}{Wischermann, K.}, \bibinfo{author}{Popp, S.},
  \bibinfo{author}{Moshir, S.}, \bibinfo{author}{Scharfetter-Kochanek, K.},
  \bibinfo{author}{Wlaschek, M.}, \bibinfo{author}{de~Gruijl, F.},
  \bibinfo{author}{Hartschuh, W.}, \bibinfo{author}{Greinert, R.},
  \bibinfo{author}{Volkmer, B.}, \bibinfo{author}{Faust, A.},
  \bibinfo{author}{Rapp, A.}, \bibinfo{author}{Schmezer, P.},
  \bibinfo{author}{Boukamp, P.}, \bibinfo{year}{2008}.
\newblock \bibinfo{title}{U{VA} radiation causes {DNA} strand breaks,
  chromosomal aberrations and tumorigenic transformation in {H}a{C}a{T} skin
  keratinocytes.}
\newblock \bibinfo{journal}{Oncogene} \bibinfo{volume}{27},
  \bibinfo{pages}{4269--4280}.
\bibitem[{Wood and Roy(2017)}]{Wood_Roy2017}
\bibinfo{author}{Wood, A.}, \bibinfo{author}{Roy, C.}, \bibinfo{year}{2017}.
\newblock \bibinfo{title}{Non-ionizing Radiation Protection}.
  \bibinfo{publisher}{John Wiley \& Sons Inc., Hoboken, USA}. chapter
  \bibinfo{chapter}{Overview: The Electromagnetic Spectrum and Nonionizing}.
\newblock pp. \bibinfo{pages}{1--10}.
\bibitem[{Wood et~al.(1999)Wood, Jimbow, Boissy, Slominski, Plonka, Slawinski,
  Wortsman and Tosk}]{Wood1999}
\bibinfo{author}{Wood, J.M.}, \bibinfo{author}{Jimbow, K.},
  \bibinfo{author}{Boissy, R.E.}, \bibinfo{author}{Slominski, A.},
  \bibinfo{author}{Plonka, P.M.}, \bibinfo{author}{Slawinski, J.},
  \bibinfo{author}{Wortsman, J.}, \bibinfo{author}{Tosk, J.},
  \bibinfo{year}{1999}.
\newblock \bibinfo{title}{What's the use of generating melanin?}
\newblock \bibinfo{journal}{Exp Dermatol} \bibinfo{volume}{8},
  \bibinfo{pages}{153--164}.
\bibitem[{Yu and Lee(2017)}]{Yu_Lee2017}
\bibinfo{author}{Yu, S.L.}, \bibinfo{author}{Lee, S.K.}, \bibinfo{year}{2017}.
\newblock \bibinfo{title}{Ultraviolet radiation: {DNA} damage, repair, and
  human disorders.}
\newblock \bibinfo{journal}{Mol Cell Toxicol} \bibinfo{volume}{13},
  \bibinfo{pages}{21--28}.

\end{thebibliography}






\newpage
\
\
\

\end{document}